\documentclass{svjour3} 
\smartqed  
\usepackage{graphicx,amsmath,enumerate}

\usepackage{bm,hyphenat,xspace}
\usepackage{graphicx,epsfig}

\newcommand {\mcu}{\mathcal{U}}

\begin{document}

\title {Properties of universal bosonic tetramers
\footnote{Special issue devoted to Critical Stability 2011}}

\author{A. Deltuva} 
\institute{ A. Deltuva  \at
Centro de F\'{\i}sica Nuclear da Universidade de Lisboa,
P-1649-003 Lisboa, Portugal \\
              \email{deltuva@cii.fc.ul.pt} } 

\date{Received: date / Accepted: date}

 \maketitle

\begin{abstract}
The system of four identical bosons is studied 
using momentum-space equations for the four-particle transition operators. 
Positions, widths and existence limits of universal unstable tetramers
are determined with high accuracy. Their effect on the 
atom-trimer and dimer-dimer scattering observables is discussed.
We show that a universal shallow tetramer intersects the atom-trimer 
threshold twice leading to resonant effects in ultracold atom-trimer 
collisions. 
\end{abstract}

\section{Introduction}

Few-particle systems with resonant interactions are universal
in the sense that their properties
are independent of the short-range interaction details.
Well known example is the Efimov effect, where 
in the unitary limit, characterized by the two-particle
scattering length $a \to \infty$,
an infinite number of weakly bound trimers 
with zero spin and positive parity ($0^+$) may exist \cite{efimov:plb}.
Furthermore, Refs.~\cite{hammer:07a,stecher:09a} predicted
the existence of two $0^+$ tetramers for each Efimov trimer.
The two lowest tetramers, i.e., the ones associated with the 
trimer ground state, are true bound states and 
 have already been studied extensively 
\cite{hammer:07a,stecher:09a,yamashita:06a}; however, 
they  may be  affected significantly by the finite-range
corrections such that even some contradictions between
Refs.\cite{hammer:07a,stecher:09a} and \cite{yamashita:06a} exist.
In contrast, all other tetramers 
  lie above the lowest particle-trimer threshold
and therefore have finite width and lifetime.
Thus, although  a number of sophisticated numerical methods 
 \cite{hammer:07a,stecher:09a,yamashita:06a,blume:00a}
is available for the four-boson bound states, not all of them
can be applied to a rigorous study of higher tetramers;
 a proper treatment of the continuum is needed.
However, in this case the technical difficulties in describing the
scattering processes involving very weakly bound dimers and trimers
in the universal regime may limit the accuracy  of the 
coordinate-space methods \cite{deltuva:ef,lazauskas:he}.
Alternative calculations using the momentum-space framework
have been recently  performed  by us for the 
atom-trimer \cite{deltuva:10c} and dimer-dimer \cite{deltuva:11b}
scattering. The description is based on the exact four-particle
Alt, Grassberger, and Sandhas 
(AGS) equations \cite{grassberger:67} for the transition operators.
The numerical technique, with some important modifications,
 is taken over from the four-nucleon scattering calculations
\cite{deltuva:07b,deltuva:08a}.
In this work it will be used to determine the universal properties
of unstable tetramers.
While their positions and limits of existence
have already been calculated using
coordinate-space methods \cite{stecher:09a,dincao:09a}, 
in our momentum-space framework we are able to achieve
the universal limit with much higher accuracy, revealing in some
cases quite drastic differences as compared to the predictions of 
Refs.~\cite{stecher:09a,dincao:09a}.
Furthermore, we obtain results for the widths of the tetramers.

In Sec.~\ref{sec:4bse} we describe the employed four-boson scattering
equations and the technical framework. 
In Sec.~\ref{sec:res} we present results for tetramer
properties and their effect on the atom-trimer and dimer-dimer
 scattering observables;
we also compare our predictions with those
 by other authors. We summarize in  Sec.~\ref{sec:sum}.

\section{Four-boson scattering equations \label{sec:4bse}}

An exact description of the four-particle scattering can be given
by the Faddeev-Yakubovsky equations \cite{yakubovsky:67}
for the wave-function components or by the equivalent
Alt, Grassberger, and Sandhas (AGS) equations \cite{grassberger:67} 
for the transition operators; the latter are 
more convenient to solve in the momentum-space framework
preferred by us. The number of independent 
transition operators (wave-function components) is significantly reduced
in the case of four identical  particles where there are only 
two distinct two-cluster partitions, one of $3+1$ type and
one of $2+2$ type. We choose those partitions to be (12,3)4 and
(12)(34) and denote them in the following by $\alpha =1$ and $2$,
respectively. The corresponding transition operators $\mcu_{\beta\alpha}$
for the system of four identical bosons obey symmetrized AGS equations 
\begin{subequations} \label{eq:U}
\begin{align}  
\mcu_{11}  = {}&  P_{34} (G_0  t  G_0)^{-1}  
 + P_{34}  U_1 G_0  t G_0  \mcu_{11} + U_2 G_0  t G_0  \mcu_{21} , 
\label{eq:U11} \\  
\mcu_{21}  = {}&  (1 + P_{34}) (G_0  t  G_0)^{-1}  
+ (1 + P_{34}) U_1 G_0  t  G_0  \mcu_{11} , \label{eq:U21} \\
\mcu_{12}  = {}&  (G_0  t  G_0)^{-1}  
 + P_{34}  U_1 G_0  t G_0  \mcu_{12} + U_2 G_0  t G_0  \mcu_{22} , 
\label{eq:U12} \\  
\mcu_{22}  = {}& (1 + P_{34}) U_1 G_0  t  G_0  \mcu_{12} . \label{eq:U22}
\end{align}
\end{subequations}
Here $G_0 = (E+i0-H_0)^{-1}$ is the free Green's function of the
four-particle system with energy $E$ and kinetic energy operator $H_0$,
the two-particle transition matrix $t$ acting within the pair (12)
is derived from the corresponding potential $v$ using the 
Lippmann-Schwinger equation
\begin{equation} \label{eq:t}
t = v + v G_0 t,
\end{equation}
 and the symmetrized operators for the 1+3 and 2+2 subsystems
are obtained from the integral equations
\begin{equation} \label{eq:U3}
U_{\alpha} =  P_\alpha G_0^{-1} + P_\alpha  t G_0  U_{\alpha}.
\end{equation}
 The employed basis states have to be symmetric 
under exchange of two particles in subsystem (12) for $3+1$ partition
and in (12) and (34) for $2+2$ partition. 
The correct symmetry of the four-boson system is ensured by the 
operators $P_{34}$, $P_1 =  P_{12}\, P_{23} + P_{13}\, P_{23}$, and
$P_2 =  P_{13}\, P_{24} $ where $P_{ab}$ is the
permutation operator of particles $a$ and $b$.

All observables for two-cluster reactions are determined by the
transition amplitudes
\begin{gather} \label{eq:ampl}
\langle \Phi_{\beta}^f| T |\Phi_{\alpha}^i \rangle  = S_{\beta\alpha}
\langle  \phi_{\beta}^f | \mcu_{\beta\alpha}| \phi_{\alpha}^{i} \rangle,
\end{gather}
obtained \cite{deltuva:07c} as on-shell matrix elements of 
the  AGS operators \eqref{eq:U}; 
the weight factors $S_{\beta\alpha}$ with values
 $S_{11} = 3$, $S_{22} = 2$, and $S_{12} = 2 S_{21} = 2\sqrt{3}$
 arise due to the symmetrization \cite{deltuva:07a}.
The matrix elements \eqref{eq:ampl} are
 calculated between the Faddeev components
\begin{equation} \label{eq:f3}
 | \phi_\alpha^{n} \rangle = G_0 \, t  P_\alpha | \phi_\alpha^{n} \rangle 
\end{equation}
of the corresponding initial/final atom-trimer or dimer-dimer  states
$| \Phi_\alpha^{n} \rangle = (1+P_\alpha)| \phi_\alpha^{n} \rangle$.

The calculation of scattering observables is done at real energies
$E = \varepsilon_{\alpha}^n + {p_\alpha^n}^2/2\mu_\alpha$
where $-\varepsilon_{\alpha}^n$ is
the binding energy of the initial $n$th state in the $\alpha$ channel,
$p_\alpha^n$ is the corresponding relative two-cluster  momentum, and 
$\mu_\alpha$ the reduced two-cluster mass. However, in the
spin/parity $0^+$ states
at complex energy values $E = E_r = -B_r - i\Gamma_r/2$,
corresponding to each unstable tetramer with energy
 $-B_r$ (relative to the four-body breakup threshold) and width $\Gamma_r$,
 the AGS transition operators \eqref{eq:U} have simple poles, i.e.,
the energy-dependence of $\mcu_{\beta\alpha}$
at $E \approx -B_r$ can be given by
\begin{equation} \label{eq:Upole}
\mcu_{\beta\alpha} =  \sum_{j=-1}^\infty
\hat{\mcu}_{\beta\alpha}^{(r,j)} (E-E_r)^{j}.
\end{equation}
The unstable bound state (UBS) pole in the complex energy plane
is located in one of the unphysical sheets that 
is adjacent to the physical sheet \cite{res_cpl}. The UBS
 therefore  affects the physical observables 
leading to resonant effects in the four-boson collisions. 
As a consequence, the properties of those
 unstable tetramers can be extracted from the behavior of
the four-boson scattering amplitudes or observables in the region
 $|E+B_r| < \Gamma_r$ where the series \eqref{eq:Upole}
is approximated very well by  few terms with $j \le 0$ or 1.

We solve the AGS equations \eqref{eq:U} in the momentum-space partial-wave
 framework with  two different types of basis states as explained
in Refs.~\cite{deltuva:ef,deltuva:07a}.
In this representation 
the  AGS equations for each total angular momentum $\mathcal{J}$
become a system of coupled  integral equations in three continuous variables,
the magnitudes of the Jacobi momenta $k_x$, $k_y$, and $k_z$
for the relative motion in the 1+1, 2+1, and 3+1 (1+1, 1+1, and 2+2)
subsystems of the 3+1 (2+2) configurations, respectively.
 Although such equations 
can be solved as done in Refs.~\cite{deltuva:07b,deltuva:08a}
for the four-nucleon scattering, the technical implementation is highly
demanding. On the other hand, in this work we are interested in the universal
properties of the four-boson system that must be independent 
of the short-range interaction details and therefore we can choose
the most convenient form of the potential. The practical solution 
simplifies considerably by using a separable two-boson potential
$v = |g\rangle \lambda \langle g|$.
In this case the AGS equations \eqref{eq:U}  can be reduced
to a system of integral equations with only 
two variables, $k_y$ and $k_z$; the details are given in 
 Ref.~\cite{deltuva:ef}.

The four-boson reactions from which we extract the tetramer properties
involve at most one dimer-dimer channel but several
(up to five in the present calculations)
atom-trimer channels with the binding
energies differing by many orders of magnitude. 
As pointed out in Ref.~\cite{deltuva:ef},
this leads to additional difficulties that are very hard
to overcome in the coordinate-space approaches 
 but can be resolved reliably in our momentum-space framework:
we discretize the  integrals using Gaussian quadrature rules 
and  use  momentum grids of correspondingly broad range;
each subsystem bound state pole of $U_\alpha$  is isolated in a different
subinterval when performing the integration over $k_z$
\cite{deltuva:07a}. 
The discretization of integrals in the AGS equations leads  to a 
system of linear algebraic equations whose solution
is described in Refs.~\cite{deltuva:ef,deltuva:07a}.

\section{Results \label{sec:res}}

The interaction model is taken over from  Ref.~\cite{deltuva:10c}, i.e.,
we use a rank-1 separable potential limited to the $l_x=0$ state 
with the form factor 
\begin{equation} \label{eq:gsep}
\langle k_x |g\rangle = [1+c_2\,(k_x/\Lambda)^2]e^{-(k_x/\Lambda)^2}
\end{equation}
and the strength $\lambda$  constrained to reproduce the given value 
of the scattering length $a$ for two particles of mass $m$.
The rank-1 potential supports at most one two-boson bound state, 
i.e., there are no deeply bound dimers.

The results will be presented as dimensionless ratios that 
are independent of the used $\Lambda$ and $m$ values 
in the universal limit. 
To demonstrate that our results are indeed independent of the details 
of the short-range interaction, we use two 
very different form factors with  $c_2 = 0$ and  $c_2=-9.17$.

\subsection{Unitary limit}

We start by presenting the results in the unitary limit
$a=\infty$ where the dimer binding energy $b_d$ vanishes but
 an infinite number of the trimers  exists with a
geometric spectrum of binding energies $ b_n = |\varepsilon_{1}^n| $,
i.e., $b_{n-1}/b_n \approx 515.035$. This number was predicted
analytically by Efimov \cite{efimov:plb} but our numerical
calculations reproduce it very well for highly excited trimers, i.e.,
for $n$ large enough such that the finite-range corrections 
become negligible. With $n \ge 4$ we achieve at least six digit 
accuracy as demonstrated in Ref.~\cite{deltuva:10c} for
both choices of the form factor (\ref{eq:gsep}).
In contrast, significant deviations were found for the ground states,
e.g., $b_0/b_1 \approx 548$ and 2126 with  $c_2=0$ and $-9.17$, respectively;
this is caused by a very different short-range behavior
of the two used models.

In our nomenclature  we characterize the tetramers by two integers $(n,k)$ 
where $n$ refers to the associated trimer and $k=1$ (2) for a deeper 
(shallower) tetramer. Our preliminary predictions for 
 the tetramer positions $B_{n,k}$  and widths $\Gamma_{n,k}$, i.e.,
their relation to the associated trimer binding energy $b_n$,
were given already in Ref.~\cite{deltuva:10c}.
Here the study of tetramer properties is improved and extended.
First we investigate the convergence
of the results with respect to the number of included partial waves
determined by the parameter $l_{\mathrm{max}}$ such that
$l_y, \, l_z \leq l_{\mathrm{max}}$. Example results for
$n=4$ and  $c_2=0$ are collected in Table~\ref{tab:res-l}. 
The convergence is quite fast but the nonzero angular momentum states
for the 2+1 and 3+1 subsystems cannot be neglected.
 Our previous results of Ref.~\cite{deltuva:10c}
obtained with $l_{\mathrm{max}}=2$ are already well converged,
the inclusion of $l_y = l_z = 3$ states yields only tiny corrections.
Furthermore, we note that the contributions of even $l_y, l_z$
are attractive while those of odd are repulsive.

\begin{table}[!]
\begin{tabular}{*{5}{c}} $l_{\mathrm{max}}$  & 
$B_{4,1}/b_4$  & $\Gamma_{4,1}/2b_4$ & 
$B_{4,2}/b_4$  & $\Gamma_{4,2}/2b_4$ 
\\  \hline
0 & 4.6754 & 0.01422 & 1.00404 & $2.93\times 10^{-4}$ \\
1 & 4.6056 & 0.01474 & 1.00215 & $2.30\times 10^{-4}$ \\
2 & 4.6108 & 0.01485 & 1.00228 & $2.38\times 10^{-4}$ \\
3 & 4.6102 & 0.01484 & 1.00227 & $2.38\times 10^{-4}$ \\
\hline
\end{tabular}
\caption{ \label{tab:res-l}
Convergence of $n=4$ tetramer properties with $l_{\mathrm{max}}$
at $a\to \infty$.
Form factor with $c_2=0$ is used.}
\end{table}

In Table~\ref{tab:res-n} we present our results
for the positions $B_{n,k}$  and widths $\Gamma_{n,k}$ of the 
tetramer pairs up to $n=5$; they were obtained with
$l_{\mathrm{max}}=3$. The ratios $B_{n,k}/b_{n}$  and  $\Gamma_{n,k}/2b_n$
for both choices of the potential form factor 
converge towards universal values
\begin{subequations} \label{eq:res}
\begin{align}
B_{n,1}/b_n = {} & 4.610(1), \\ 
\Gamma_{n,1}/2b_n = {} & 0.01483(1), \\
B_{n,2}/b_n = {} & 1.00227(1), \\ 
\Gamma_{n,2}/2b_n = {} & 2.38(1)\times 10^{-4}
\end{align}
\end{subequations}
as $n$ increases. However, significant potential-dependent
deviations due to finite-range effects
can be seen for $B_{n,k}/b_{n}$  and  $\Gamma_{n,k}/2b_n$
at $n \le 1$ and  $n \le 2$, respectively.
With $c_2=-9.17$, where the $n=0$ trimer is a non-Efimov-like state
\cite{deltuva:10c}, the $(0,2)$ tetramer is even absent.
Including a strong repulsive three-body force 
would decrease the binding energies and increase the size of the states
and thereby speedup the $n$-convergence \cite{stecher:09a} 
but the ground state calculations ($n=0$) would be insufficient anyway
since $n=0$ doesn't account for the inelastic collisions and finite width.
This also explains why the convergence for $\Gamma_{n,k}$ is slower
than for  $B_{n,k}$.

\begin{table}[!]
\begin{tabular}{*{5}{c}} $n$ &
$B_{n,1}/b_{n}$  & $\Gamma_{n,1}/2b_n$ & $B_{n,2}/b_n$  & $\Gamma_{n,2}/2b_n$ 
\\  \hline
0 &   5.6402  &  & 1.04185 &  \\
1 & 4.5169 & 0.03363 & 1.00105 & $3.82\times 10^{-4}$ \\
2 & 4.6035 & 0.01366 & 1.00216 & $2.14\times 10^{-4}$ \\
3 & 4.6098 & 0.01471 & 1.00226 & $2.36\times 10^{-4}$ \\
4 & 4.6102 & 0.01484 & 1.00227 & $2.38\times 10^{-4}$ \\
5 & 4.6102 & 0.01483 & 1.00227 & $2.38\times 10^{-4}$ \\
\hline
0 & 3.2192  &  &  &  \\
1 & 4.9923 & 0.01360 & 1.00996 & $4.18\times 10^{-4}$ \\
2 & 4.6108 & 0.02084 & 1.00227 & $3.34\times 10^{-4}$ \\
3 & 4.6098 & 0.01493 & 1.00226 & $2.39\times 10^{-4}$ \\
4 & 4.6102 & 0.01483 & 1.00227 & $2.38\times 10^{-4}$ \\
5 & 4.6102 & 0.01483 & 1.00227 & $2.38\times 10^{-4}$ \\
\hline
\end{tabular}
\caption{ \label{tab:res-n} 
Tetramer positions and widths in the unitary limit.
Results obtained with $c_2=0$ ($c_2=-9.17$) in Eq.~(\ref{eq:gsep})
are given in the top (bottom) part.}
\end{table}

\begin{figure}[!]
\includegraphics[scale=0.45]{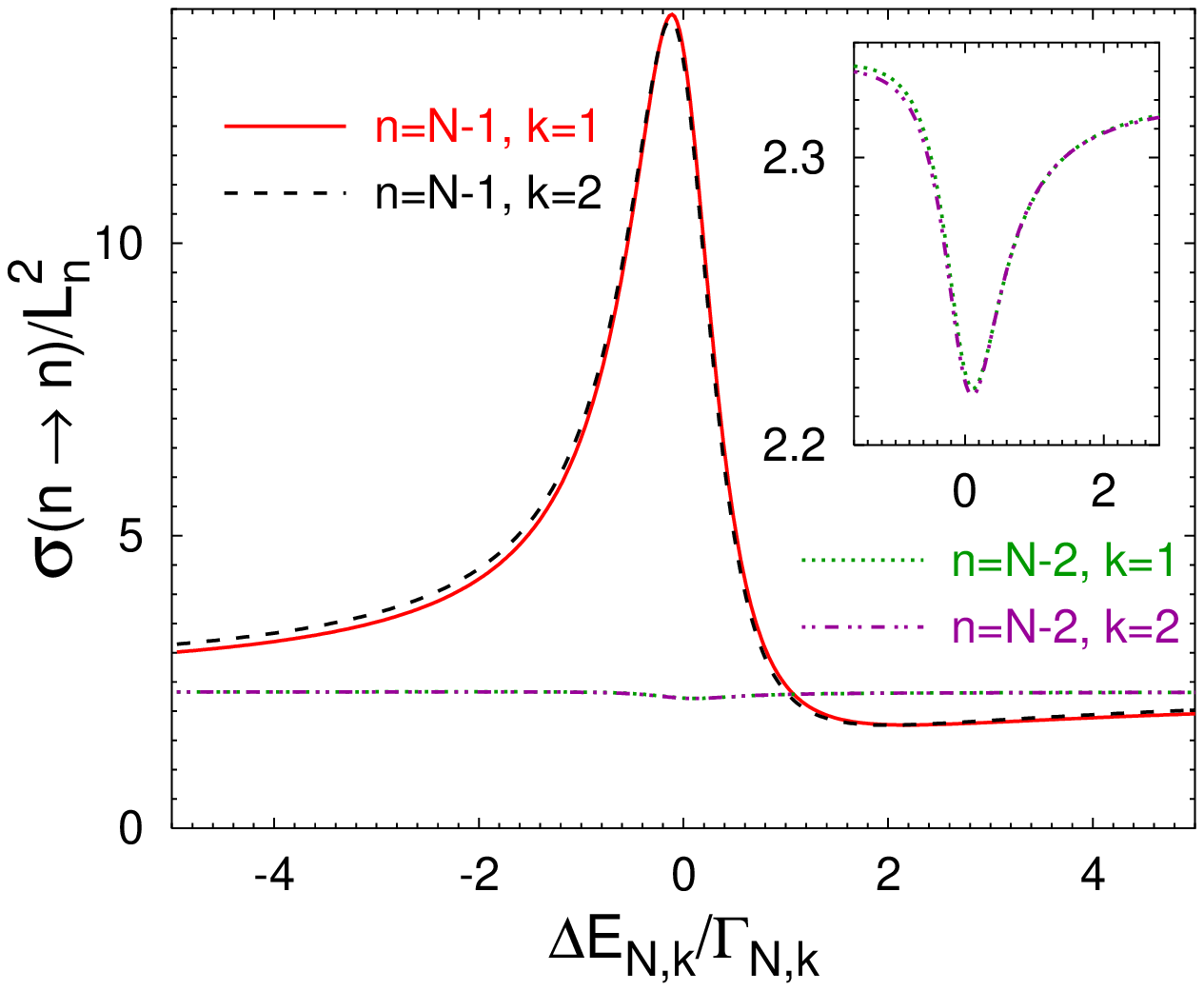} \quad
\includegraphics[scale=0.45]{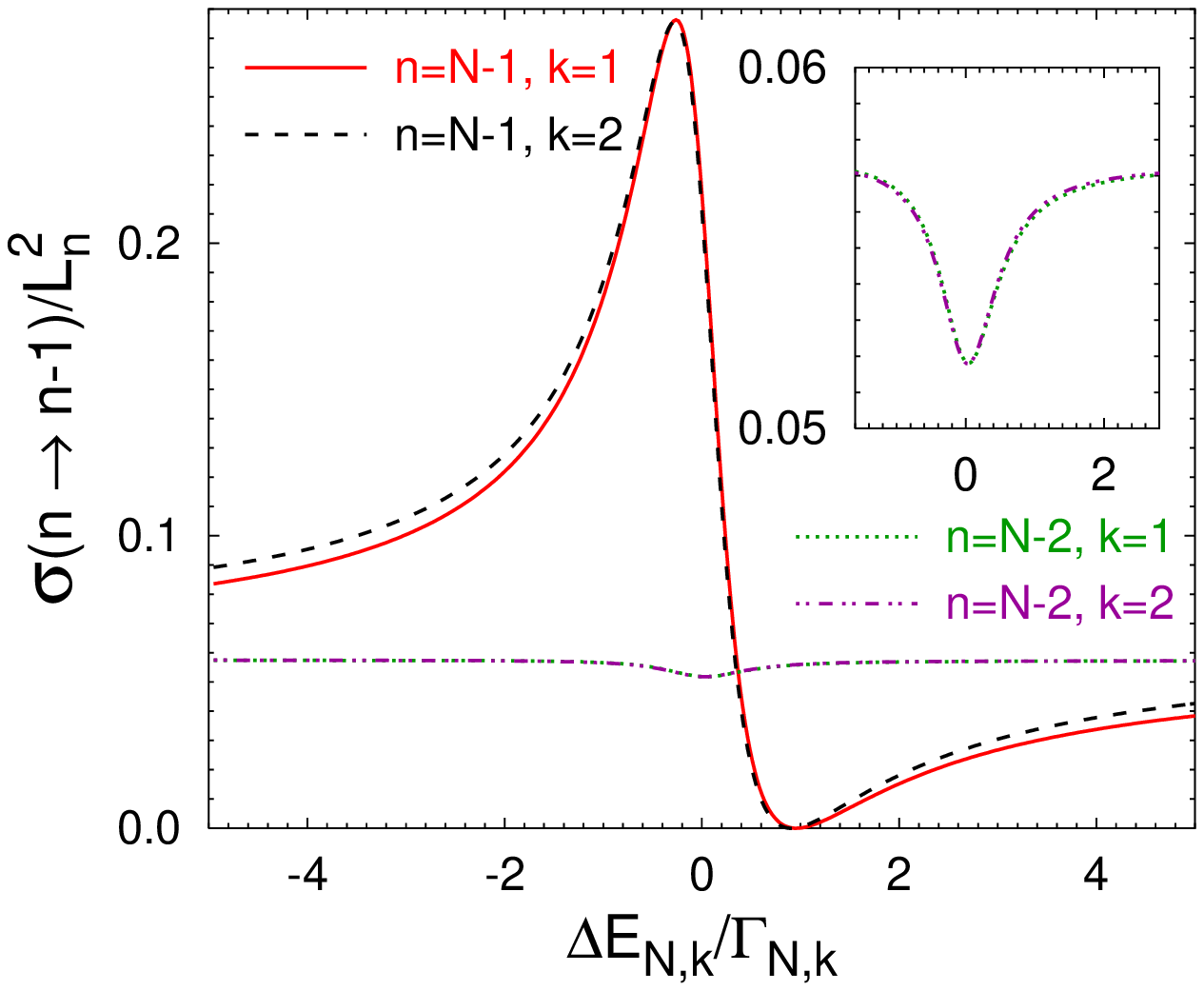}
\caption{\label{fig:rs-cs} (Color online)
Elastic and inelastic cross sections 
for the atom scattering from the $n$th  trimer
in the vicinity of the $(N,k)$th tetramer.}
\end{figure}

Next we study the effect of unstable tetramers on the
elastic and inelastic cross sections  $\sigma(n\to n')$
in atom-trimer collisions;  $n$ and $n'$ characterize the trimer state
in the initial and final channel, respectively.
To form dimensionless ratios for each trimer we introduce
the length scale  $L_n = \hbar / \sqrt{2\mu_1 b_n}$.
As already found in  Ref.~\cite{deltuva:10c}, for sufficiently large $n$
and $n'$ the ratios $\sigma(n\to n')/L_n^2$ 
depend only on $n-n'$ but not on the employed potential.
Furthermore, for the  inelastic cross sections ($n'<n$) an additional
relation ${\sigma(n\to n')} /{\sigma(n\to n'-1)} \approx 43.7$
was established \cite{deltuva:10c}.
Thus, in the universal limit the atom scattering from the $n$th trimer can be 
fully characterized by only two cross sections, the elastic one $\sigma(n\to n)$
and the leading inelastic one $\sigma(n\to n-1)$. 
In the case of $\sigma(n\to n)$  the atom-trimer $P$- and $D$-wave contributions
calculated in Ref.~\cite{deltuva:10c} have to be added; they
are negligible for  $\sigma(n\to n-1)$.
In Fig.~\ref{fig:rs-cs} we study the behavior of  the elastic and inelastic
atom-trimer cross sections in the vicinity of the 
$(N,k)$th tetramer; we use $c_2=0$ and $N=5$ such that the 
finite-range effects are negligible.
 We use the energy variable $\Delta E_{N,k} = E + B_{N,k}$ that
measures the distance to the tetramer position.
Despite very different tetramer widths,
the behavior of the cross section as function of
$\Delta E_{N,k}/\Gamma_{N,k}$ is very similar for  $k=1$ and 2.
Although $\mcu_{11}$ has pole in all open channels $n<N$,
the elastic and inelastic cross sections $\sigma(n\to n)$ and 
$\sigma(n\to n-1)$ have characteristic resonance peaks only in
the case of $n=N-1$ where they increase by a factor of 5.
For $n \le N-2$ a minimum is seen close to $\Delta E_{N,k} = 0$
 which becomes less and less
pronounced as the difference $N-n$ increases; we therefore show
only  $n = N-2$ results where in the minimum the elastic (inelastic)
cross section is decreased by 5\% (10\%). 
In accordance with this behavior the phase shift (not shown here) 
increases by $180^{\circ}$ only for $n = N-1$ while  local minima take place
for  $n \le N-2$.

\subsection{Tetramers between four-atom and dimer-dimer thresholds}

\begin{figure}[!]
\includegraphics[scale=0.58]{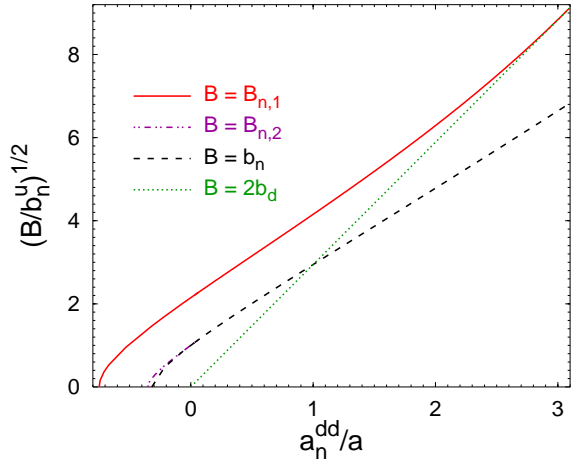}
\caption{\label{fig:Ba} (Color online)
Tetramer, trimer and dimer binding energies as functions of 
the two-boson scattering length.}
\end{figure}

\begin{figure}[!]
\includegraphics[scale=0.58]{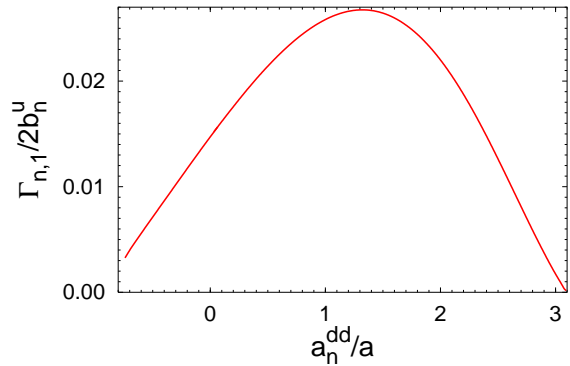}
\caption{\label{fig:G1} (Color online)
The width of the deeper tetramer as a function of
the two-boson scattering length.}
\end{figure}

Trimers and the associated  tetramers exist in a certain
regime of  large finite $|a|$.
In Fig.~\ref{fig:Ba} we show the tetramer positions  $B_{n,k}$  
as functions of $a$; we include also the binding energies in the 
atom-trimer and dimer-dimer channels, $b_n$ and $2b_d$.
As a reference point for $a$ we choose the intersection of the dimer-dimer and 
the $n$th atom-trimer thresholds, i.e., $b_n = 2 b_d$ at $a = a_n^{dd}$.
All the binding energies are normalized by the 
$n$th trimer binding energy in the unitary limit $b_n^u$.
In Fig.~\ref{fig:Ba} we show $B_{n,2}$ only at  $a < 0$; at $a>0$
the shallow tetramer lies very close the  atom-trimer threshold and
exhibits a nontrivial behavior that will be presented 
separately in the next subsection.
The $a$-evolution of the width  $\Gamma_{n,1}$ of the deeper tetramer
is shown in Fig.~\ref{fig:G1}.

\begin{figure}[!]
\includegraphics[scale=0.58]{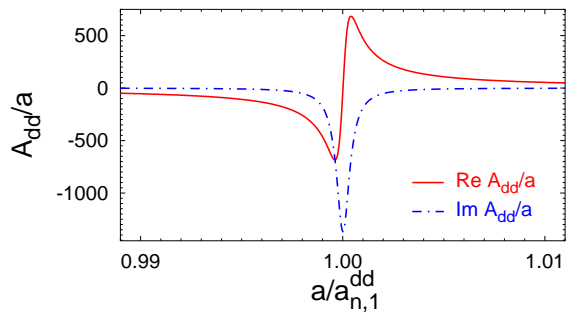}
\caption{\label{fig:add1} (Color online)
Dimer-dimer scattering length as a function of the atom-atom 
scattering length in the vicinity of the $k=1$ tetramer intersection
with the dimer-dimer threshold.}
\end{figure}

As can be seen in  Fig.~\ref{fig:Ba},
on the side of negative $a$ the trimers (tetramers)
emerge at the three  (four)  free atom threshold with zero energy.
We denote by $a_n^0$ and $a_{n,k}^0$ the specific negative values of $a$ 
where  $b_n=0$ and  $B_{n,k}=0$, respectively.
In an ultracold atomic gases  these $a$ values would correspond to a
resonant enhancement of the three or four-atom recombination process
 \cite{stecher:09a,ferlaino:09a}. 
On the side of positive $a$ the trimers decay via the atom-dimer threshold,
i.e., $b_n = b_d$ at  $a = a_n^{d}$;
this situation is outside the range of Fig.~\ref{fig:Ba} since
$a_{n}^{dd}/ a_{n}^{d} =  6.789(1)$ \cite{deltuva:11b}.
The tetramers decay via the dimer-dimer threshold, i.e.,
 $B_{n,k} \approx 2b_d$ and  $\Gamma_{n,k} = 0$ at  $a = a_{n,k}^{dd}$,
leading to a resonant behavior of the dimer-dimer scattering length
$A_{dd}$ shown for $k=1$ in Fig.~\ref{fig:add1}.
The consequence of this phenomenon in an ultracold gas of dimers 
is a resonant enhancement of the trimer creation and dimer-dimer relaxation
processes \cite{ferlaino:08a,dincao:09a,deltuva:11b},
 the zero-temperature rate being 
$ \beta_{dd}^0 = -(8\pi \hbar/m) \mathrm{Im} A_{dd}$.

All those special values of $a$ are related in a universal way provided
$n$ is sufficiently large. With the present interaction models 
 the universal limit is reached with high accuracy
at $n=4$ as can be seen in numerous examples for tetramer properties
given in Table~\ref{tab:res-n} and for various atom-trimer scattering 
observables in Ref.~\cite{deltuva:10c}. In particular,
the convergence for the tetramer intersections with the dimer-dimer threshold 
is demonstrated in Ref.~\cite{deltuva:11b}, yielding the ratios 
\begin{subequations} \label{eq:a-dd}
\begin{align}
a_{n,1}^{dd}/ a_{n}^{dd} & =  0.3235(1), \\
a_{n,2}^{dd}/ a_{n}^{dd} & =  0.99947(2),
\end{align}
\end{subequations}
where the uncertainties are estimated by comparing the predictions obtained
 with different $n$ and $c_2$. 
For the intersection of the tetramers with the four free atom threshold
we get
\begin{subequations} \label{eq:a-0}
\begin{align}
a_{n,1}^{0}/ a_{n}^{0} & =  0.4254(2), \\
a_{n,2}^{0}/ a_{n}^{0} & =  0.9125(2).
\end{align}
\end{subequations}
Furthermore, semi-analytical results \cite{braaten:rev} are available
for some quantities of the three-boson nature, namely,
$ a_{n}^{d} \sqrt{m b_n^u} = 0.0707645086901 $
and $- a_{n}^{0} \sqrt{m b_n^u} = 1.56(5)$ for $n \to \infty$.
They agree well with our numerical predictions
\begin{subequations} \label{eq:a-3}
\begin{align}
 a_{n}^{d} \sqrt{m b_n^u}  & = 0.07076(1), \\
 -a_{n}^{0} \sqrt{m b_n^u}  & = 1.5077(1)
\end{align}
\end{subequations}
thereby confirming their reliability.

\subsection{Shallow tetramer} 

\begin{figure}[!]
\includegraphics[scale=0.64]{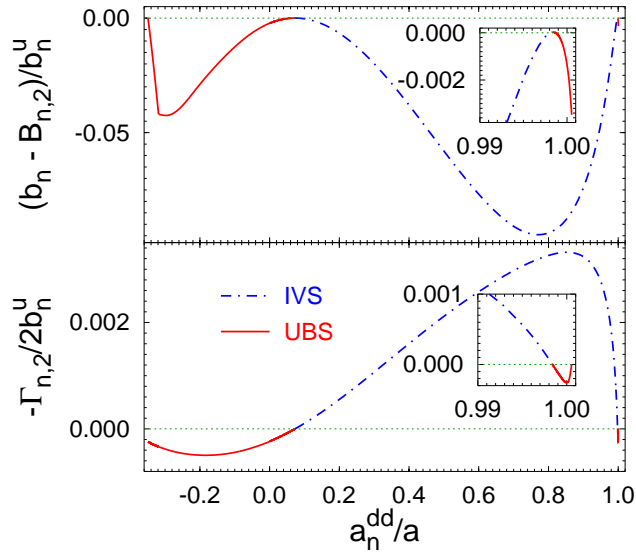} 
\caption{ 
Position of the shallow tetramer relative to the 
atom-trimer threshold (top) and its width (bottom)
as functions of the two-boson scattering length $a$.
We used $b_n = 0$ at $a_{n}^{dd}/a < a_{n}^{dd}/a_n^0 \approx -0.3186$ 
where the $n$th trimer doesn't exist.}
\label{fig:BG2}
\end{figure}

 Since the shallow  tetramer lies very close to the
associated atom-trimer threshold, we use a different representation
to show the $a$-evolution of its position, namely, we consider
its relative distance to the atom-trimer threshold
$(b_n-B_{n,2})/  b_n^u$. Together with the width it is presented
in Fig.~\ref{fig:BG2} in the whole region of its existence.
For both tetramers the widths $\Gamma_{n,k}$ remain finite at 
the respective $a = a_{n,k}^0$ where $B_{n,k}$ vanish;
the $k=1$ case is shown in  Fig.~\ref{fig:G1}.
The shallow tetramer detaches most from the atom-trimer threshold
around $a \approx a_n^0$ where $b_n$ almost vanishes.
Most remarkably, at two special positive values of $a = a_n^{v,j}$
the shallow tetramer intersects the atom-trimer threshold, i.e.,
when moving away from the unitary limit it first
decays at $a_n^{v,1}$,  then reappears at  $a_n^{v,2}$, and finally
decays via the dimer-dimer threshold at $a = a_{n,2}^{dd}$.
In other words, the shallow tetramer in a particular regime 
$a \in (a_n^{v,2},a_n^{v,1})$  becomes an inelastic virtual state (IVS)
 \cite{res_cpl}
with $\Gamma_{n,2} < 0$  instead of an UBS with positive width. 
The intersection points $a_n^{v,j}$ correspond to  $\Gamma_{n,2} = 0$
and are universal, i.e.,
\begin{subequations} \label{eq:a-ivs}
\begin{align}
a_n^{v,1}/a_n^{dd} & =  13.75(5) \\
a_n^{v,2}/a_n^{dd} & = 1.0016(1) . 
\end{align}
\end{subequations}
In the vicinity of $a = a_n^{v,j}$ the 
tetramer position  $B_{n,2} < b_n$ while otherwise  $B_{n,2} > b_n$.
Thus, it may seem that the tetramer IVS is mostly below the 
atom-trimer threshold. However, one has to keep in mind the changed sign 
of $\Gamma_{n,2}$ for IVS that implies the change of the energy sheet.
 The IVS corresponds to the pole of the transition operators $\mcu_{\beta\alpha}$
in the complex energy plane 
on one of the nonphysical sheets that is, in contrast to the one of UBS,
more distant from the physical sheet \cite{res_cpl}. For this reason the
IVS affects the physical scattering observables in a completely
different way as compared to UBS: the elastic and inelastic cross sections 
around $E \approx -B_{n,2}$ for IVS, i.e., for $a \in (a_n^{v,2},a_n^{v,1})$
 show no resonant peaks that were seen in Fig.~\ref{fig:rs-cs} for the UBS. 
Thus, the parameters of the IVS cannot be extracted using
Eq.~\eqref{eq:Upole}. However, an approximate procedure based 
on the atom-trimer scattering length and effective range parameter
as described in Ref.~\cite{deltuva:11a} was applied to obtain the
IVS results in Fig.~\ref{fig:BG2}.  
The IVS pole only affects the  physical observables when it is
located extremely close to the atom-trimer threshold.
In that case  the cross sections and phase shift have a cusp exactly at 
the atom-trimer threshold, i.e., at $E = -b_n$, but the cusp disappears rapidly 
with increasing $(B_{n,2}-b_n)$ and $-\Gamma_{n,2}$;
an examples can be found in Ref.~\cite{deltuva:11a}.
We note that the tetramers become IVS also after crossing the
dimer-dimer threshold, i.e., at $a < a_{n,k}^{dd}$. 

The most prominent effect of the tetramer UBS-IVS conversions 
through the atom-trimer (dimer-dimer) threshold 
 is a resonant enhancement of the atom-trimer
scattering length $A_n$  around $a = a_n^{v,j}$ 
(dimer-dimer scattering length $A_{dd}$  around $a = a_{n,k}^{dd}$);
thus, at the corresponding $a$ values $A_n$ 
exhibits qualitatively the same behavior as shown for $A_{dd}$ 
in Fig.~\ref{fig:add1}. The regime around  $a_n^{v,j}$ is not 
yet explored experimentally. However,
in an ultracold mixture of atoms and excited trimers tuning
the atom-atom scattering length to the values close to $a_n^{v,j}$
would lead to a resonantly increased rate of the atom-trimer relaxation
whose zero-temperature limit is $ \beta_{n}^0 = -(4\pi \hbar/\mu_1) \mathrm{Im} A_{n}$.
Of course, in real experiments the  resonance positions may
deviate from the universal values (\ref{eq:a-ivs}) 
due to finite-range effects.

\subsection{Comparison with previous works \label{sec:comp}}

A number of numerical techniques is available for the four-boson bound 
state calculations, both in the momentum \cite{hammer:07a,yamashita:06a}
and coordinate space \cite{stecher:09a,blume:00a}.
 Some of them \cite{stecher:09a},  neglecting the finite width
of the higher tetramers,  were used to find 
their energies  and threshold intersection points.
The results of Refs.~\cite{stecher:09a,dincao:09a} for
$B_{n,k}/b_{n}$ at unitarity,  $a_{n,k}^{0}/a_{n}^{0}$, and 
$a_{n,k}^{dd}/a_{n}^{dd}$ agree with ours within few percents.
Those calculations
\cite{stecher:09a,dincao:09a}  were limited to $n \le 1$ or 2
where the finite-range effects could not be entirely neglected
and the convergence with $n$ was not better than few percents;
this is consistent with our findings as can be seen in Table~\ref{tab:res-n},
albeit with different potentials. We demonstrated that
higher $n$ are needed to achieve the universal limit accurately;
this is technically very demanding, especially
in the coordinate-space framework.

However, in contrast to our work,
Ref.~\cite{stecher:09a} has not predicted the shallow 
tetramer intersections with the atom-trimer threshold at $a_n^{v,j}$. 
It appears that the results of Refs.~\cite{stecher:09a,dincao:09a} are
quite poorly converged for the fine-scale quantities $(B_{n,2}/b_n-1)$
at unitarity and $(1-a_{n,2}^{dd}/ a_{n}^{dd})$.
For the former our five best-converged results 
are 0.00227 within 0.5\%  according to Table~\ref{tab:res-n} while
the three best-converged results of Ref.~\cite{stecher:09a} are
0.006, 0.03, and 0.001. For  $(1-a_{n,2}^{dd}/ a_{n}^{dd})$
the prediction of Ref.~\cite{dincao:09a}, 0.019, overestimates our
well-converged result 0.00053 by a factor
of 35. Both these deviations indicate that  Refs.~\cite{stecher:09a,dincao:09a}
strongly overestimate the distance of the shallow tetramer
from the atom-trimer threshold  such that they never cross each other.
We note that the  nonuniversal $(0,2)$ tetramer of our work is
also bound considerably stronger than the universal ones and therefore
doesn't intersect the atom-trimer threshold.

\section{Summary \label{sec:sum}}

We studied universal bosonic tetramers that are unstable bound states in 
the continuum and strongly affect collisions  in the four-boson system.
We extracted the tetramer properties such as positions, widths, and limits of
existence from the behavior of the atom-trimer and dimer-dimer scattering 
observables. These collision processes were described using
exact four-particle scattering equations for the transition operators
that were solved in the momentum-space framework with high precision.
A rigorous treatment of the four-boson continuum enabled us to determine 
the widths of the tetramers that were out of reach in previos works. 
Furthermore, we accurately achieved the universal limit by considering reactions
involving high excited trimers where the finite-range effects are negligible.
In this respect our results are much better converged than those of previous
works \cite{stecher:09a,dincao:09a} where only the tetramer positions and existence 
limits have been calculated. While the agreement between our predictions
and those of Refs.~\cite{stecher:09a,dincao:09a} is reasonable for the more
tightly bound tetramer, there are drastic differences for the shallow one. 
We demonstrate that changing the two-boson scattering 
length the shallow tetramer intersects the atom-trimer threshold twice
and in a special regime becomes an inelastic virtual state;
these UBS-IVS conversions lead to resonant effects in ultracold
atom-trimer collisions.


\begin{thebibliography}{10}

\bibitem{efimov:plb}
Efimov, V.: Energy levels arising from resonant two-body forces in a three-body
  system. Phys. Lett. B {\bf 33},  563  (1970).

\bibitem{braaten:rev}
Braaten, E., Hammer, H.-W.: Universality in few-body systems with large
  scattering length. Phys. Rep. {\bf 428},  259  (2006).

\bibitem{hammer:07a}
Hammer, H.~W., Platter, L.: Universal properties of the four-body system with
  large scattering length. Eur. Phys. J. A {\bf 32},  113  (2007).

\bibitem{stecher:09a}
von Stecher, J., D'Incao, J.~P., Greene, C.~H.: Signatures of universal
  four-body phenomena and their relation to the Efimov effect. Nature Phys.
  {\bf 5},  417  (2009).

\bibitem{ferlaino:08a}
Ferlaino, F., Knoop, S., Mark, M., Berninger, M., Sch\"obel, H., N\"agerl,
  H.-C., Grimm, R.: Collisions between Tunable Halo Dimers: Exploring an
  Elementary Four-Body Process with Identical Bosons. Phys. Rev. Lett. {\bf
  101},  023201  (2008).

\bibitem{ferlaino:09a}
Ferlaino, F., Knoop, S., Berninger, M., Harm, W., D'Incao, J.~P., N\"agerl,
  H.-C., Grimm, R.: Evidence for Universal Four-Body States Tied to an Efimov
  Trimer. Phys. Rev. Lett. {\bf 102},  140401  (2009).

\bibitem{pollack:09a}
Pollack, S.~E., Dries, D., Hulet, R.~G.: Universality in Three- and Four-Body
  Bound States of Ultracold Atoms. Science {\bf 326},  1683  (2009).

\bibitem{deltuva:10c}
Deltuva, A.: Efimov physics in bosonic atom-trimer scattering. Phys.~Rev.~A
  {\bf 82},  040701(R)  (2010).

\bibitem{deltuva:11b}
Deltuva, A.: Universality in bosonic dimer-dimer scattering. Phys.~Rev.~A {\bf
  84},  022703  (2011).

\bibitem{yamashita:06a}
Yamashita, M.~T., Tomio, L., Delfino, A., Frederico, T.: . Europhys. Lett. {\bf
  75},  555  (2006).

\bibitem{blume:00a}
Blume, D., Greene, C.~H.: Monte Carlo hyperspherical description of helium
  cluster excited states. J. Chem. Phys. {\bf 112},  8053  (2000).

\bibitem{deltuva:ef}
Deltuva, A., Lazauskas, R., Platter, L.: Universality in few-body scattering.
  Few-Body Syst. {\bf 51},  235  (2011).

\bibitem{lazauskas:he}
Lazauskas, R., Carbonell, J.: Description of He$_4$ tetramer bound and
  scattering states. Phys. Rev. A {\bf 73},  062717  (2006).

\bibitem{grassberger:67}
Grassberger, P., Sandhas, W.: Systematical treatment of the non-relativistic
  n-particle scattering problem. Nucl. Phys. {\bf B2},  181  (1967); E. O.
  Alt, P. Grassberger, and W. Sandhas, JINR report No. E4-6688 (1972).

\bibitem{deltuva:07b}
Deltuva, A., Fonseca, A.~C.: Four-body calculation of
  proton-{}${}^3\mathrm{He}$ scattering. Phys.~Rev.~Lett. {\bf 98},  162502
  (2007).

\bibitem{deltuva:08a}
Deltuva, A., Fonseca, A.~C., Sauer, P.~U.: Four-nucleon system with
  $\Delta$-isobar excitation. Phys.~Lett.~B {\bf 660},  471  (2008).

\bibitem{dincao:09a}
D'Incao, J.~P., von Stecher, J., Greene, C.~H.: Universal Four-Boson States in
  Ultracold Molecular Gases: Resonant Effects in Dimer-Dimer Collisions. Phys.
  Rev. Lett. {\bf 103},  033004  (2009).

\bibitem{yakubovsky:67}
Yakubovsky, O.~A.: On the integral equations in the theory of N particle
  scattering. Yad. Fiz. {\bf 5},  1312  (1967) [Sov. J. Nucl. Phys. {\bf 5},
  937 (1967)].

\bibitem{deltuva:07c}
Deltuva, A., Fonseca, A.~C.: Ab initio four-body calculation of
  $n$-${}^3\mathrm{He}$, $p$-${}^3\mathrm{H}$, and $d$-$d$ scattering.
  Phys.~Rev.~C {\bf 76},  021001(R)  (2007).

\bibitem{deltuva:07a}
Deltuva, A., Fonseca, A.~C.: Four-nucleon scattering: Ab initio calculations in
  momentum space. Phys.~Rev.~C {\bf 75},  014005  (2007).

\bibitem{res_cpl}
Badalyan, A.~M., Kok, L.~P., Polikarpov, M.~I., Simonov, Y.~A.: Resonances in
  coupled channels in nuclear and particle physics. Phys. Rep. {\bf 82},  31
  (1982).

\bibitem{deltuva:11a}
Deltuva, A.: Shallow Efimov tetramer as inelastic virtual state and resonant
  enhancement of the atom-trimer relaxation. EPL {\bf 95},  43002  (2011).

\end{thebibliography}

\end{document}